\documentclass[conference]{IEEEtran}
\IEEEoverridecommandlockouts

\IEEEoverridecommandlockouts                              

\usepackage[english]{babel}

\usepackage{amsmath}
\usepackage{graphicx}
\usepackage[colorinlistoftodos]{todonotes}
\usepackage{soul} 
\usepackage{dblfloatfix} 
\usepackage{balance} 

\begin{document}

\title{Security Implications of Fog Computing on the Internet of Things}
\author{
\IEEEauthorblockN{Ismail Butun$^{1*}$, Alparslan Sari$^{2*}$, and Patrik \"{O}sterberg$^{1}$}
\IEEEauthorblockA{\textit{$^{1}$Department of 
Information Systems and Technology,} 
\textit{Mid Sweden University,}
Sundsvall, Sweden \\
\textit{$^{2}$Department of Computer Engineering, University of Delaware,} 
Newark, Delaware, USA \\
e-mails:\textit{ismail.butun@miun.se, asari@udel.edu, patrik.osterberg@miun.se}}
\thanks{$^{*}$Corresponding authors and have equal contribution.}
\thanks{-This work was supported by SMART Project, EU Regional Fund (grant number 20201010).}
}



\maketitle

\begin{abstract} Recently, the use of IoT devices and sensors has been rapidly increased which also caused data generation (information and logs), bandwidth usage, and related phenomena to be increased. To our best knowledge, a standard definition for the integration of \textit{fog computing} with IoT is emerging now. This integration will bring many opportunities for the researchers, especially while building cyber-security related solutions.  In this study, we surveyed about the integration of fog computing with IoT and its implications. Our goal was to find out and emphasize problems, specifically security related problems that arise with the employment of fog computing by IoT. According to our findings, although this integration seems to be non-trivial and complicated, it has more benefits than the implications.
\end{abstract}


\section{Introduction}
Internet of Things (IoT) is having its hipe now, as the Internet had its hipe two decades ago. IoT market is expected to grow from more than 15 billion devices three years ago to more than 75 billion in 2025 \cite{dataeconomy}. IoT needs a strong technological foundation for its rapid development and acceptance from the scientific community. Hence, the fog computing is a very strong candidate to provide this foundation for IoT. By providing several advantages, fog computing is expected to be one of the main backbone pillars of the IoT in terms of computational support.

\begin{figure}[b]
\centering
\includegraphics[width=0.5\textwidth]{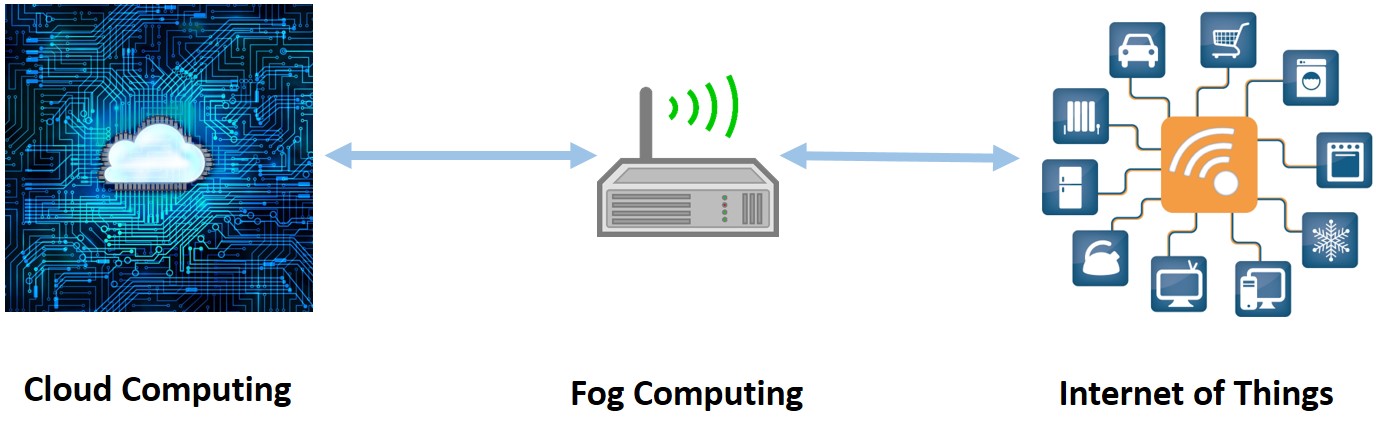}
\caption{\label{fig1} Fog computing proposed as a gateway in between cloud computing and IoT.}
\end{figure}

As shown in Fig.~\ref{fig1}, from a conceptual point of view, we are predicting fog computing to serve as an intermediate level of service for seamlessly handshaking the protocols of cloud computing and IoT. This will bring many benefits: 1) Cloud computing servers are super fast in contrast to the IoT devices. Fog computing devices will provide an interface between the two far set of devices. 2) This intermediate layer of fog computing will allow several fixes (such as patch updates, etc.) to be done easier. Instead of making changes on IoT devices, software updates can be pushed on to the fog device(s). 3) Fog computing will bring all the advantages of edge-computing, such as the agility, scalability, decentralization, etc.

As a centralized resource out of users’ control, the cloud represents every possible opportunity to violate privacy. Unfortunately, privacy has become a luxury today, a situation that will be exacerbated in the IoT \cite{zhang2015cloud}. Therefore, a remedy is needed to enhance the privacy needs of the users in these services and fog computing is a strong candidate to provide this.

Fog computing actually is a tool for cloud-based services (CBS) that can be thought of as an interface in between the real end-devices and the rest of the CBS. CBS  offers three service models, namely Infrastructure as a Service (IaaS), Platform as a Service (PaaS) and Software as a Service (SaaS) \cite{butun2015anomaly}. We are projecting that fog computing paradigm will act as an interface for these CBS service models so that intended services can be used by the front-end users seamlessly and promptly.

The \textit{Security Plane} for CBS proposed by Butun \textit{et al.} \cite{butun2015anomaly} was intended to be used for the front-end IoT devices and to be an interface to the cloud. After the proposal of fog computing, this \textit{Security Plane} kind of solution is highly implementable. Therefore, we think of fog computing to provide extra services such as security to the edge of the cloud for the CBS. For example, the usage of fog computing would bring benefits to the Intrusion Detection Systems (IDS) that are devised for IoT. Hence early detection is important to stop ill effects of intrusions, fog computing would bring early detection opportunities to IDS algorithms working on IoT.

Fog computing brings three immediate advantages over cloud computing: $1)$ Enhanced service quality to mobile users. $2)$ Enhanced efficiency to the network. $3)$ Enhanced location awareness. Among these benefits, the major benefit of fog computing over the cloud is that the support for location awareness which might be very useful for the applications that are employing location based services (LBS) \cite{A9:LuanG0XS15}.

\begin{figure*}
\centering
\includegraphics[width=1.0\textwidth]{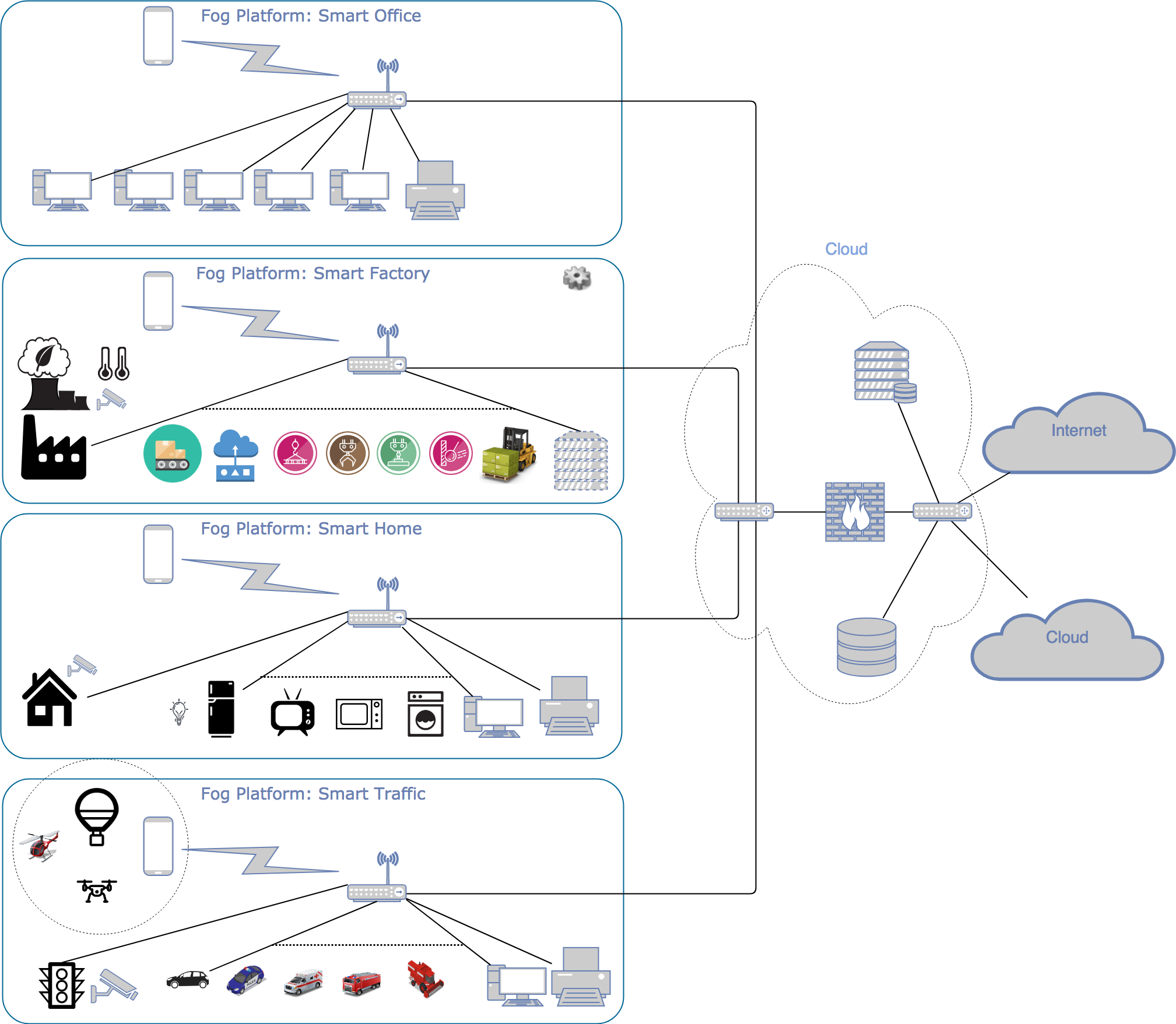}
\caption{\label{fig2} An illustration of four different possible fog computing applications with IoT: Smart Office, Smart Factory, Smart Home and Smart Traffic.}
\end{figure*}

\section{Background}
Large-scale IoT deployments created situations which cloud computing could not handle efficiently and effectively. For instance, applications which require low latency while processing the data on the edge of the network. In real life, a massive amount of data is being collected by IoT from many different sensors in various environments such as factory production lines, vehicles, machines, elevators etc. or individual purposes such as smart home systems, hobby related sensors, etc. 

These sensing devices have different characteristics and features. They are connected to each other via hardwire or WiFi. Large-scale device deployments in heterogeneous environments bring management issues. Hence, intelligent communications approaches are needed in which efficiency and robustness are prioritized.

Using a cloud network to stream data and analyze data has its limitations such as bandwidth consumption and communication costs. If the user data are sensitive, securing the data is another important issue. The data are important for auditing purpose or controlling the assets to improve efficiency or preventing disasters etc.

The data analysis could be done on site by running the software at local stations. The cloud would be used as storing the analysis result for historical and audit purposes. The data aggregation will reduce the bandwidth and also bandwidth related cost.

\begin{table*}
\caption{Comparison of cloud and fog computing concepts.}
\label{table1}
\small
\begin{center}
\begin{tabular}{|l||c|c|}
\hline
\textbf{Feature} & \textbf{Cloud computing} & \textbf{Fog computing}\\
\hline
\hline
Access & Wired or wireless& Wireless\\
\hline
Access to the service & Through server & At the edge device\\
\hline
Availability & Mostly available & Mostly volatile\\
\hline
Content distributed to & Edge device & Anywhere\\
\hline
Content generator & Man made & Sensor made\\
\hline
Content generation at & Central server & Edge device\\
\hline
Control & Centralized & Distributed\\
\hline
Latency & High & Minor\\
\hline
Location of resources (i.e. processing and storage) & Center& Edge\\
\hline
Mobility & Not supported & Supported\\
\hline
Number of users & Millions & Billions \\
\hline 
Virtual infrastructure location & Enterprise server & User devices\\
\hline
\end{tabular}
\end{center}
\end{table*}

Fig.~\ref{fig2} presents various possible application fields of fog computing: Smart Office concept can be an example of the generic relation of IoT devices and fog computing. Smart Factory is an example of industrial IoT (IIoT) and fog computing application. There could be many IoT devices, sensors (temperature, pressure etc.), electric actuators or other control devices could be involved. Smart Home concept is emerging with IoT devices and home appliances such as TV, washing machine, dryer, refrigerator etc. as they are getting smarter and intelligent. In Smart Traffic example, data collection on site and immediately analyzing and processing data on the edge may help in fast decision making locally, instead of sending data to a central location. For instance, in case of an emergency, the traffic lights can be controlled to open a way for emergency vehicles such as fire trucks and ambulances based on local IoT devices. In these four different scenarios, the common idea is that the devices generate a massive amount of data and may need to collaborate with each other and to take critical decisions reducing the delay. Hence, an agile response is important and the philosophy of fog computing may help to overcome bandwidth and latency related problems in this manner.

Because of introducing agile response nearby the edge components, we are expecting fast implementation and business growth of fog computing for future IoT applications such as smart-traffic and smart-factories. Thereby, the integration will not remain in just IoT but expand to industrial IoT (IIoT) and further other areas. This will impose its own challenges to IIoT \cite{forsstrom2018challenges}, as well as bringing benefits.

IoT and fog computing can be helpful in designing ``smart'' things such as smart home, smart traffic lights, smart cities, etc. For instance, the sensors in a smart traffic system can detect accidents or sense the road conditions due to weather or some other factors and inform the drivers. A traffic jam can be regulated by a smart traffic system.

In recent years, due to the usage of IoT and other sensors, the data generated by end-devices increased massively. The question is where/when/how should these data be analyzed? In cloud-centric design, IoT devices generate data and send them to the cloud (operates as a central server) for storage and analyses. However, in fog computing, the data is analyzed on the edge stations and just necessary results are being sent to the cloud.

Fog computing concept is recently introduced by CISCO \cite{computing2015internet}, which is a new vision that enables IoT devices to run on the edge of the network. According to Bonomi \textit{et al.} \cite{A1:Bonomi:2012}, ``Fog Computing" is not an alternative for ``Cloud Computing". Fog extends the cloud computing and complements the cloud computing with the concept of smart devices which can work on the edge of the network. According to CISCOs vision, fog computing has following characteristics: 1- Low latency, 2- Location awareness, 3- Geographical distribution, 4- Mobility, 5- Very large number of nodes, 6- A predominant role of wireless access, 7- Streaming and real-time applications, 8- Heterogeneity \cite{computing2015internet}.

OpenFog Consortium \cite{openfogconsortium} is defining the standards of the fog computing with different committees and work-groups. The founding members are Arm, Cisco, Dell, Intel, Microsoft and Princeton University. The focus is to create and promote an open reference architecture for fog computing to solve challenges such as bandwidth, latency, etc. in various areas like AI, IoT, industrial machinery, Robotics, etc. According to OpenFog Consortium, the key pillars of the fog architecture are security, scalability, openness, autonomy, reliability, availability, serviceability, agility, hierarchy, and programmability. Fog computing is helping to the IoT, 5G and AI related systems which need some special unique properties such as security (trusted transactions), cognition (objective awareness), agility (scalable), latency (real-time processing), and efficiency (utilizing unused resources). According to OpenFog, the benefits of using fog computing are; low latency, business agility, security, real-time analytics, reduced cost, less bandwidth usage.

Table \ref{table1} presents fog computing and cloud computing concepts in a comparative way \cite{stallings2018computer}. As can be seen, fog computing presents more agile and rapid response when compared to cloud computing, henceforth represents a strong candidate as a technological solution for future IoT and IIoT based implementations. Fog computing would be a preferable approach with various IoT designs and applications such as Smart Home, Smart Traffic (Transportation and Connected vehicles etc.), Smart Grid, Industrial Automation and integration with IIoT, Smart Health-care Systems, etc.



The benefits of fog computing for IoT (and IIoT) can be summarized as follows:
\begin{itemize}
\item\textbf{Reducing cost:} The data will be processed on edge rather then cloud
\item\textbf{Reducing the delay:} Critical applications require low latency to interpret the data and to take a decision. The cloud computing is not suitable to serve this task.
\item\textbf{Agile response:} Real-time applications may benefit from fog computing concept to gain speed during analysis or decision making phase.
\item\textbf{Increased security:} With fog computing, service providers can filter out sensitive personally identifiable information (PII) and process it locally, sending the non-sensitive information to the cloud for further processing \cite{microsoft}.
\end{itemize}






\section{Practical application scenarios of Fog Computing usage in IoT}
In this section, we introduced a subset of possible attack scenarios from real-life and discussed possible related mitigation methods.

Coin, bill and card-based systems suffer from multiple attack vectors. Asking the following question ``How to hack a vending machine?" on Google returns thousands and thousands of results. Whether the hacks work or not does not really mean anything, the information of possible attack vectors poses a great threat to existing vending machine domain since one attack vector may work with an old (vulnerable) model. In early 80-90s, many coin-operated machines including but not limited to vending machines, arcade machines, public phones etc. suffered from simple Coin-on-a-String trick \cite{tvropes}. Later, this problem is mitigated with the one-way ratchet. Coin or card systems can be bypassed by attacker \cite{youtube:laundry1}, the machine can be tampered \cite{youtube:laundry2}. Paper bills tampered with plastic tape or other materials to get them back from machine after the credit is earned or attacker tries to machine spit the bill back after it is credited.

Even the vending machines in one of the most secured places on the world can be hacked, such as the ones in Central Intelligence Agency (CIA) facilities. According to BuzzFeed News, a group of CIA contractors exploited a vulnerability in the vending machines electronic payment systems to buy snacks at no cost \cite{buzzfeed}. The total loss is around \$3,314, but the most shocking fact is even vending machines in CIA are vulnerable and the unexpected can happen. Apparently, attackers disconnected the FreedomPay network cable connected to vending machines to exploit an ``Availability" issue which resulted in the machine permitting purchases made by unfunded FreedomPay cards. Attackers are later on identified by agency's surveillance cameras.

The extreme example above from real-life has shown us that we may more way to go in securing vending machines kind of payment systems. Luckily, now fog computing IoT technology enables us to implement more secure and agile payment systems as follows:

\subsection{Smart Laundromats}
In the real life, there have been many cases of laundromat system hacking reported \cite{reddit}. 
Accordingly, it can be stated that none of the laundromat systems (whether token based, magnetic card based, or even smart card based) that are being used today is secure. One of the main reasons is that the machine usage is not controlled via the server but at the machines. When a machine is hacked, or a token (card) is hacked, there is no mechanism to check the validity (whether it is authentic or not) of the transaction. Therefore, we project real-time  (or close to real-time) usage of authentication mechanism that is governed by a central server (Authentication Management) and served to the edge devices (laundromats) with the adoption of fog computing.

\subsection{Smart Vending Machines}
As an application of fog supported-IoT to vending machines, a vending machine can report missing items to the vendor so that they can be shipped on time. Besides, auto status check report can be generated and can be sent to the vendor for maintenance purposes.

Smart vending machines mitigates the security problems but much more can be achieved from a ``smart'' machine. For instance, detailed sales reports can be generated for the vendor. A particular product can be tracked throughout the year, and the vendor can get detailed insight about the sales of a product or user buying habits like which product sales are best, when, where, etc. Product-A sales are best in February, user-1 buys product-B on Monday morning but buys product-C on Tuesday evening, etc. 

\begin{table*}
\small 
\caption{Possible security considerations when the Fog Computing gateway is compromised}
\label{table2}
\begin{center}
\begin{tabular}{|c||c|c|}
\hline
\hline
\textbf{Feature} & \textbf{Risk on IoT network} & \textbf{Impact on Cloud}\\
\hline
Access control & Moderate & Minor\\
\hline
Authentication & Moderate/Significant & Minor/Major\\
\hline
Availability & Significant& Minor\\
\hline
Confidentiality & Moderate& Minor\\
\hline
Integrity & Minimal & None/Minor\\
\hline
Privacy & Significant & Minor/Major\\
\hline
\end{tabular}
\end{center}
\end{table*}

\subsection{Smart Chip Card Systems}
Smart Chip Card Systems are relatively secure than coin or paper bill based systems. The threat model for these systems is unencrypted or not well-encrypted cards. With a card reader/writer, the value in the card can be easily updated if the necessary preemptive steps are not taken. Another attack vector would be, an attacker may clone the card and get unlimited credit if there is not a mechanism to do validation (checking transaction with a server etc.) If IoT is embedded with the device in the production phase, it is unlikely that it will be physically tampered by an attacker easily (assuming human security details are overseeing the machine). A possible solution is designing a fog computing-based system which runs a micro-service to validate the transactions. Smart Chip Card can be an identifier for a legal user, once it is used in a device, the micro-service on fog computing system validates the user and transaction or escalates the situation and asks mobile application based authentication. For instance, if the card is used in a different location (same apartment complex but different building etc.) this can be reported as an instance.

Another distinct advantage of using fog computing-based IoT is the reduction of possible operational cost. For example, end-devices can connect to the smart grid and negotiate with the grid on the unit electric price. On a queue based approach, they can operate their tasks when the electric price is cheap (during the daytime for instance).


\section{Projections of Fog Computing usage in IoT and its Security Implications}
According to the scientific projections, fog computing is expected to be one of the main backbone structure of IoT in the near future \cite{dataeconomy, openfogconsortium,microsoft}. Inevitably, there will be implications of this integration. 

In ideal conditions, extra introduced components are desired to bring no further burden on the overall operation of the existing system. However, this is not the fact in real life scenarios. Sometimes they bring an extra load (e.g. processing and memory storage), and sometimes (preferably) decrease existing load. The same is valid for security features. 

Security implications of using fog computing for IoT systems is shown in Table \ref{table2}. There are six features (most important ones being CIA, i.e. confidentiality, integrity and availability) that we considered in the case of a failure (capture) in defending the fog computing gateway (FCG):
\begin{enumerate}
\item \textit{Access Control:} FCG devices provide a gateway between IoT network and cloud. However, the functioning is like providing field data from the IoT to cloud and command messages from the cloud to IoT. There is no way of accessing databases of the cloud from FCG devices, on the other hand, it is possible to command and conquer all IoT devices connected to a rouge FCG device.
\item \textit{Authentication:} Depending on the authentication algorithm design, if several layered authentication methodology (one at FCG, one at the cloud, etc.) is used then this might be a very secure solution. Any compromise would only affect limited number of IoT devices.  Otherwise, if all authentication operation is left to the FCG devices, this might create problems when the FCG devices get compromised.
\item \textit{Availability:} Depending on the critical position of the FCG, we expect a significant impact on the availability of the IoT resources if the communications are blocked. However, this will not affect the cloud side marginally.
\item \textit{Confidentiality:} Confidentiality of the data at FCG has a moderate impact on IoT network, whilst has a minor impact on the cloud (all the devices connected to FCG gets affected, however, the rest of the IoT network and the cloud remains safe operation). 
\item \textit{Integrity:} Depending on the communications scheme, we expect minor (if end-to-end encryption is not employed) to none (with encryption)  effect of FCG on the integrity of the messages.
\item \textit{Privacy:} As in any service, privacy of the users in IoT is critical and any leakage through FCG devices might have serious consequences. This will affect all users that are using the IoT though that hacked FCG device. However, the rest of the data (resulting from other FCGs) at the cloud will still be secure and private. 
\end{enumerate}

If these computers (gateways) are installed in a few numbers, they may constitute a single point of failure. Since, if an attacker manages to harm this gateway, all the communication between cloud and IoT would be blocked. Therefore, we suggest several gateways installed network architectural implementation.

As discussed above and also listed in Table \ref{table1}, compromising the fog computing gateway, will have an impact on cloud computing layer from minor to critical levels; and the risk on IoT network will be from minimal to significant levels, depending on the security feature that is observed. Therefore, in some applications, the security of fog computing gateway might be very important. Hardware and/or software security precautions for the fog computing gateway should be considered while deciding security provisioning for the overall network.

The fog computing layer can be leveraged by security services as a proxy server with the firewall and/or IDS capability. Most of the attacks can be prevented by the firewall. Any harmful intrusion attempt can be detected by IDS and mitigated at the fog computers before it can reach to main servers on the cloud.

To improve privacy, a wiser suggested solution would be keeping the private data on the edge while sending the just necessary data to the cloud. 

\section{Future Remarks and Conclusions}
The proliferation of IoT devices in our surrounding environment is indispensable and this will be possible due to the dominant usage of fog computing as a backbone supporting architecture.  Throughout this article, we have discussed the implications of using fog computing as a backbone architecture for IoT, especially from cyber security point of view. 

According to our findings, we have stated that usage of fog computing for cloud-based IoT systems might have several benefits; in terms of cost, QoS and more importantly, security.

\balance 



\bibliographystyle{IEEEtran} 
\bibliography{main} 

\end{document}